\documentclass[aps,prb,twocolumn,floatfix,showpacs]{revtex4}
\usepackage{amsfonts}
\usepackage{amsmath}
\usepackage{amssymb}
\usepackage{graphicx}
\usepackage{color}
\usepackage{bm}

\begin{document}

\title{In-plane magnetic field anisotropy of the FFLO state in layered
superconductors}
\author{M. D. Croitoru$^{1}$}
\author{M. Houzet$^2$}
\author{A. I. Buzdin$^1$}
\affiliation{$^1$Universit\'{e} Bordeaux I, LOMA, UMR 5798, F-33400
Talence, France}
\affiliation{$^2$SPSMS, UMR-E 9001,
CEA-INAC/UJF-Grenoble 1, F-38054 Grenoble, France}

\begin{abstract}
There are strong experimental evidences of the
Fulde-Ferrell-Larkin-Ovchinnikov (FFLO) state formation in layered organic
superconductors in parallel magnetic field. We study theoretically the
interplay between the orbital effect and the FFLO modulation in this case
and demonstrate that the in-plane critical field anisotropy drastically
changes at the transition to the FFLO state. The very peculiar
angular dependence of the superconducting onset temperature
which is predicted may serve
for unambiguous identification of the FFLO modulation. The obtained results
permit us to suggest the modulated phase stabilization as the origin of the
magnetic-field angle dependence of the onset of superconductivity
experimentally observed in (TMTSF)$_{2}$ClO$_{4}$ organic conductors.
\end{abstract}

\date{\today }

\maketitle


Layered superconductors exposed to an external magnetic field aligned
parallel to their conducting layers have been in the focus of theoretical
and experimental investigations due to their remarkable anisotropic
properties\cite{buzdin00,lebe00} favorable to the formation of the spatially
modulated phase, known as the Fulde-Ferrell-Larkin-Ovchinnikov (FFLO) state.%
\cite{lark01,fuld} In particular, in the family of organic layered
superconductors (TMTSF)$_{2}$X, where anion X is PF$_{6}$, ClO$_{4}$, etc.,
very large upper critical fields, which exceed the Pauli paramagnetic limit,
were reported.\cite{lee01,lee02,lee03,oh,yone01,yone02} In layered
conductors the orbital motion of electrons is mostly restricted to the
conducting crystal planes when hopping between adjacent layers is small.
Thus the magnetic field applied parallel to the conducting planes causes
only small diamagnetic currents and the orbital depairing is strongly
weakened. Therefore, spin-singlet superconductivity is mainly limited by the
Zeeman energy (Pauli spin polarization) of the quasiparticles. In contrast,
the Pauli effect is negligible for a spin-triplet pairing because in this
case Cooper pairs gain Zeeman energy without loosing the condensation
energy. The question concerning the singlet or triplet symmetry of the
superconducting order parameter in layered organic conductors is a current
topic of debate. Indeed, the Nuclear Magnetic Relaxation (NMR) experiments
with (TMTSF)$_{2}$PF$_{6}$ salts below $T_{c}$ and under pressure showed the
absence of the Knight shift, thus supporting the triplet scenario of pairing,%
\cite{lee04} while the $^{77}$Se NMR Knight shift in a recent
experiment with (TMTSF)$_{2}$ClO$_{4}$ revealed a decrease in spin
susceptibility consistent with singlet pairing.\cite{shin02}
$^{13}$C NMR measurements with
$\kappa$-(BEDT-TTF)$_{2}$Cu(NCS)$_{4}$ also evidenced for a
Zeeman-driven transition within the superconducting state and
stabilization of FFLO phase.\cite{wrig} For the singlet
superconductivity the FFLO phase can be a candidate for the
enhancement of the upper critical field.\cite{lark01,fuld}
{\color{black}{Note that in the compound (TMTSF)$_{2}$ClO$_{4}$ the
substantial anisotropy within the conducting
$\mathbf{a}-\mathbf{b}^{\ast }$ plane is present. When magnetic
field is aligned along the high conductivity $\mathbf{a}$-axis the
orbital currents are strongly quenched, which favors the FFLO phase
appearance.\cite{buzdin06,lebe02} Interestingly, for a magnetic
field applied along the $\mathbf{b}^{\ast }$-axis, the 3D
$\rightarrow$ 2D dimensional crossover occurs in the high field
regime and the coexistence of the hidden reentrant and FFLO phases
can emerge.\cite{lebe03}}}

Recently the in-plane angular dependence of upper critical field,
$H_{c2}$, of the organic superconductor (TMTSF)$_{2}$ClO$_{4}$ has
been measured for wide temperature intervals.\cite{yone01} The
observed upturn of the $H_{c2}$ curve at low temperatures has often
been discussed in connection with the possibility of the FFLO state
formation.\cite{shim01,buzdin02} In addition as shown in Ref.
\onlinecite{yone02} the superconducting phase in high magnetic field
is more strongly suppressed by
impurities than that in low field, as expected in the FFLO scenario.\cite%
{asla} Furthermore, an unusual in-plane anisotropy of $H_{c2}$ in the
high-field regime was observed, which was again interpreted as an evidence
of the FFLO state stabilization. This argument is based on the prediction of
a very peculiar in-plane angular dependence of the FFLO critical field due
to the orbital effects in thin superconducting films.\cite{buzdin05}
Motivated by these experimental findings we investigate in this work the
influence of the spatially modulated superconducting phase on the in-plane
anisotropy of the upper critical field in layered superconductors with
\textit{s}-wave pairing.

To describe the layered superconductors we consider a system of layers in
\textit{xy}-plane, stacked along the \textit{z}%
-axis. The single-electron spectrum is approximated by
\begin{equation}
E_{\mathbf{p}}=\frac{p_{x}^{2}}{2m_{x}}+\frac{p_{y}^{2}}{2m_{y}}+t\cos
\left( p_{z}d\right) ,
\end{equation}%
where $\mathbf{p}=(p_{x},p_{y},p_{z})$ is the electron momentum. The
in-plane motion is described within the effective mass approximation
while the tight-binding approximation is used to describe the motion
along the \textit{z}-direction. The corrugation of the Fermi surface
due to the coupling between adjacent layers (interlayer distance
$d$) is assumed to be small, i.e. $t\ll T_{c0}$, but sufficiently
large to make the mean-field treatment valid, $\left\vert \ln
\left( T_{c0}/t\right) \right\vert T_{c0}/E_{F}\ll 1$.\cite{tsu}
Here $T_{c0}$ is the critical temperature of the system at zero
magnetic field and $E_{F}$ is the Fermi energy.
We choose a gauge for which the vector potential $\mathbf{A}=\mathbf{H}%
\times \mathbf{r}$ [$\mathbf{r}=(x,y,0)$ is a coordinate in \textit{xy}%
-plane], i.e. $A_{z}=-xH\sin \alpha +yH\cos \alpha $, where $\alpha
$ is the angle between the applied field, with amplitude $H$, and
\textit{x}-axis. As was demonstrated in Ref. \onlinecite{brison} the
anisotropic model with effective masses can be reduced to the
isotropic one by a scaling transformation and corresponding
renormalization of the magnetic field. {\color{black}{Therefore, in
the pure Pauli regime, the orientation of the FFLO modulation
vector, $\mathbf{q}$, is arbitrary in the case of an elliptical
Fermi surface. However any deviation of the Fermi surface from the
ellipticity fixes the direction of the modulation.
\cite{buzdin05,shim03,denisov} Hereinafter we assume for the sake of
simplicity that these deviations from ellipticity are small and
their role is just to pin the direction of the vector $\mathbf{q}$,
which is supposed to be along the \textit{x}-axis.}} Performing this
scaling transformation, we will thus consider from now on an
isotropic in-plane spectrum, with mass $m=m_{x}$, and a magnetic field $%
\mathbf{H}=H[({m_{x}}/{m_{y}})^{1/2}\cos \alpha ,\sin \alpha,0]$. Taking into
account that the system is near the second-order phase transition, the
linearized Eilenberger equation on the anomalous Green function $f_{\omega }(%
\mathbf{n},\mathbf{r},p_{z})$ describing layered superconducting systems
acquires the form (for positive Matsubara frequency $\omega $
at temperature $T$)\cite{kop}
\begin{equation}
\lbrack \omega +ih+\frac{1}{2}\mathbf{v}_{F}.\bm{\nabla }+2it\sin
(p_{z}d)\sin (\mathbf{Q.r})]f_{\omega }(\mathbf{n},\mathbf{r},p_{z})=\Delta (%
\mathbf{r}).  \label{HH01}
\end{equation}%
Here $h=\mu _{B}H$ is the Zeeman energy,
$\mathbf{v}_{F}=v_{F}\mathbf{n}$ is the in-plane Fermi velocity, and
$\mathbf{Q}=(\pi dH/\phi _{0})[-\sin \alpha ,(m_{x}/m_{y})^{1/2}\cos
\alpha ,0]$ with $\phi _{0}=\pi c/e$. The order parameter is defined
self-consistently as
\begin{equation}
\frac{1}{\lambda }\Delta \left( \mathbf{r}\right) =2\pi T\operatorname{Re}%
\sum\limits_{\omega >0}\left\langle f_{\omega }\left( \mathbf{n},\mathbf{r}%
,p_{z}\right) \right\rangle ,  \label{Delta}
\end{equation}%
where $\lambda $ is the BCS pairing constant and the brackets denote
averaging over $p_{z}$ and $\mathbf{n}$. Here we considered a
layered superconductor in the clean limit, meaning that the in-plane
mean free path is much larger than the corresponding coherence
length, $\xi_0= v_F/(2\pi T_{c0})$. The upper critical
field corresponds to the values of $H$ for which the system of Eqs. (\ref%
{HH01}) and (\ref{Delta}) can be solved.

The solution of the Eilenberger equation (\ref{HH01}) can be chosen
without loss of generality as a Bloch function
\begin{equation}
f_{\omega }\left( \mathbf{n},\mathbf{r},p_{z}\right) =e^{i\mathbf{qr}%
}\sum\limits_{m}e^{im\mathbf{Q.r}}f_{m}\left( \omega ,\mathbf{n}%
,p_{z}\right) .  \label{HH02}
\end{equation}%
Equation (\ref{HH02}) takes into account the possibility for the formation
of the pairing state $\left( \mathbf{k}+\frac{\mathbf{q}}{2},\uparrow ;-%
\mathbf{k}+\frac{\mathbf{q}}{2},\downarrow \right) $ with finite
center-of-mass momentum. At the same time, the order parameter can
be expanded as
\begin{equation}
\Delta ( \mathbf{r}) =e^{i\mathbf{qr}}\sum\limits_{m}e^{i2m\mathbf{Q.r}%
}\Delta _{2m}.  \label{HH04}
\end{equation}%

It is known\cite{sarma} that in the absence of orbital effect, the FFLO
state only appears at $T<T^{\ast }\simeq 0.56T_{c0}$ and $H>H^{\ast }\simeq
1.06T_{c0}/\mu _{B}$, where $(T^{\ast },H^{\ast })$ is the tricritical
point. Therefore, the order of the magnitude of the magnetic field required
to observe the FFLO state can be found from the relation $\mu _{B}H\sim
T_{c0}$. Taking this into account one obtains $v_{F}Q\sim v_{F}edT_{c0}/\mu
_{B}c
\sim (d/a)T_{c0}$, where $a$
is the unit cell in the \textit{xy}-plane. Therefore $v_{F}Q\gtrsim T_{c0}$.
Due to the assumption $t\ll T_{c0}\lesssim v_{F}Q$ one has $\sqrt{tT_{c0}}%
\ll v_{F}Q$. This condition allows us to retain only the terms up to the
first harmonics in Eqs. (\ref{HH02}) and (\ref{HH04}) [since we will retain
only the terms up to $\left( t/T_{c0}\right) ^{2}$ in the final
expressions]. Substituting Eqs. (\ref{HH02}) and (\ref{HH04}) into Eq. (\ref%
{HH01}) one gets
\begin{eqnarray}
L(\mathbf{q})f_{0}+\tilde{t}f_{-1}-\tilde{t}f_{1} &=&\Delta _{0},  \notag \\
L(\mathbf{q\pm Q})f_{\pm 1}\pm \tilde{t}f_{0} &=&0,  \label{HH05}
\end{eqnarray}%
where $L(\mathbf{q})=\omega +ih+i\mathbf{v}_{F}.\mathbf{q}/2$ and $\tilde{t}%
=t\sin (p_{z}d)$. If one neglects the Zeeman term these equations
readily describe
the reentrant phase predicted by Lebed,\cite{lebe01} with critical temperature $T_{c0}$
at fields $H\gg \phi_0/(d\xi_{0})$. While keeping the terms
up to the second harmonics within the same procedure would yield the
Lawrence-Doniach equation.\cite{lebe01} Inserting the solution of Eqs. (\ref%
{HH05}) into the self-consistency equation (\ref{Delta}), keeping only terms
up to the second order in $t/T_{c0}$, and subtracting it with a similar
equation relating $\lambda $ with $T_{c0}$, we obtain
\begin{align}
\ln \frac{T_{c0}}{T}=& 2\pi T\operatorname{Re}\sum\limits_{\omega >0}\left\{ \frac{1%
}{\omega }-\left\langle \frac{1}{L(\mathbf{q})}\right\rangle \right.
\label{HH06} \\
& \left. +
\left\langle \frac{\tilde{t}^{2}}{L^{2}(\mathbf{q%
})}\left(
\frac{1}{L(\mathbf{q}+ \mathbf{Q})}+
\frac{1}{L(\mathbf{q}- \mathbf{Q})}
\right)
\right\rangle \right\} .  \notag
\end{align}%
This equation defines the temperature dependence of the upper critical
magnetic field $H_{c2}^{{}}$\ in layered superconductors, when both the
paramagnetic and orbital effects are accounted for.

\begin{figure*}
\resizebox{1.75\columnwidth}{!}{\rotatebox{0}{
\includegraphics{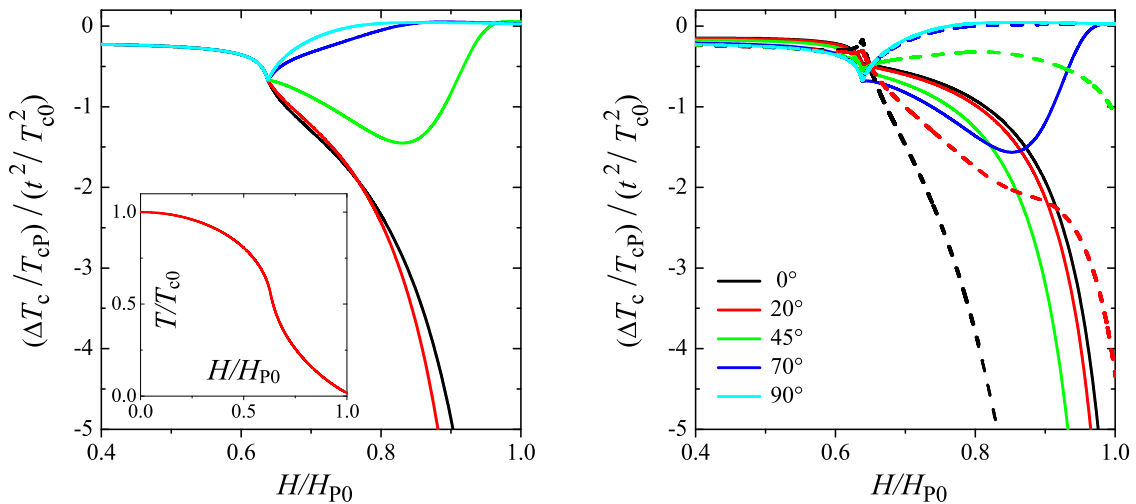}}}
\caption{Reduced critical temperature (with respect to the critical
temperature in the pure paramagnetic limit) as a function of
in-plane magnetic field in a layered superconductor, for several
angles $\alpha$ between the field and $x$-axis (equivalently FFLO
modulation vector at $H>H^\ast$) Left panel: isotropic regime with
$m_{x}=m_{y}$. [Inset: $(H,T)$-phase diagram in the pure
paramagnetic limit.] Right panel: anisotropic regime. Solid lines
are for $m_{x}=10m_{y}$; dashed lines are for $m_{x}=0.1m_{y}$.
The calculations are performed for 
Fermi
velocity $v_{F}=2.10^{7}~\mathrm{cm.s}^{-1}$,\protect\cite{lebe02} and
interlayer distance $d=1.3~\mathrm{nm}$. \protect\cite{yone01}}
\label{fig.01}
\end{figure*}

In the limit $t \ll T_{c}$ the magnitude of the FFLO modulation
vector can be calculated by neglecting the orbital part in Eq.
(\ref{HH06}). When making average over the Fermi surface one gets
the equation
\begin{eqnarray}\label{HH07}
\ln \frac{T_{c0}}{T} =F( \tilde{h},\tilde{q})
&\equiv &
\sum_{n=0}^\infty
(
[{n+1/ 2}]^{-1}
\\
&&
-[(n+1/2+i\tilde {h})^2+{v_F^2\tilde q^2}/{4}]^{-1/2}
),
\notag
\end{eqnarray}%
with reduced variables $\tilde{h}= h/2\pi T$ and $\tilde {q}%
= q/2\pi T$, which gives rise to a
FFLO vector $\mathbf{q}$ with magnitude that maximizes the
upper critical field, thus defining $T_{cP}(H)$ and $q_P(H)$, in the pure Pauli limit.%
\cite{bula02}
Finally averaging Eq. (\ref{HH06}) over the Fermi surface one
obtains the equation for the onset of superconductivity in layered
superconductors, $T_c(H)$, in the presence of both Zeeman and
orbital effects: 
\begin{eqnarray}
\frac{T_{cP}-T_c}{T_{c}} &=&\frac{1} {1-\tilde{h} \partial F(
\tilde{h},\tilde{q}) /\partial \tilde{h} }
\label{HH08} \\
&&
\left.
\times 2\pi T\operatorname{Re}\sum\limits_{\omega > 0,\pm }
\left\langle
\frac{\tilde{t}^{2}}
{L^{2}(\mathbf{q})L(\mathbf{q}\pm
\mathbf{Q})}
\right\rangle
\right|_{
      T=T_{cP}
        } . \notag
\end{eqnarray}

The summation over the Matsubara frequencies is performed numerically. We
used $N=10^{4}$ terms in the summation and this number suffices for
convergency at $T/T_{c0}> 10^{-2}$. Fig. \ref{fig.01} shows
the variation of the normalized correction of the transition temperature, $%
\Delta T_{c}=T_{c}-T_{cP}$, as a function of reduced strength of the
magnetic field, $H/H_{P0}$ and angle $\alpha $ (between $\mathbf{H%
}$ and $x$-axis). Here $H_{P0}=\Delta _{0}/\mu _{B}$ is the critical
magnetic field at $T=0$ in Pauli limited two-dimensional
superconductors.\cite{bula02} [The ($H,T$)-phase diagram in this
regime is given in the inset of left panel.] The left panel
describes the isotropic situation{\color{black}{, typical for layered
quasi-2D compounds,\cite{singlet02} while the right panel exhibits
results obtained for the highly anisotropic in-plane Fermi surface
of layered conductors, exhibiting quasi-1D character.\cite{lebe04}
We consider two opposite mass anisotropies. When $m_{x}=10m_{y}$
$\mathbf{q}$ is along the heavy mass direction, while in the case of
$m_{x}=0.1m_{y}$ it is along the light mass direction.}} As it was
intuitively expected the orbital effects reduce the superconducting
onset temperature, $\Delta T_c<0$. While increasing the applied
magnetic field, $\Delta T_{c}$ first decreases in most cases until
the tricritical point, $H^{\ast }$, is reached and the curve of
$\Delta T_{c}$ exhibits a kink. At $H>H^{\ast }$ the function
$\Delta T_{c}(H)$ strongly depends on the
in-plane effective mass anisotropy and angle $\alpha$. For $\alpha $ close to $90%
{{}^\circ}%
$, $\Delta T_{c}$ exhibits an upturn and $T_{c}$ approaches the
paramagnetic limit, $T_{cP}$, when $H$ increases. In contrast, for small $\alpha
$ an increase of the magnetic field leads to a decrease of $\Delta T_{c}$.
For intermediate angles, $\Delta T_{c}$ can be a non-monotonic
function of the field strength.
In the isotropic case and for $H/H_{P0}\gtrsim 0.75$ the largest
correction to the onset temperature $\left\vert \Delta T_{c}^{{}}\left(
\alpha \right) \right\vert $\ occurs at $\alpha \approx 20%
{{}^\circ}%
$. For $m_{x}/m_{y}=10$ and $H/H_{P0}\gtrsim 0.8$, $\left\vert \Delta
T_{c}^{\max }\right\vert $ is at angles close to $\alpha \approx 45%
{{}^\circ}%
$, while for $m_{x}/m_{y}=0.1$ and $H/H_{P0}\gtrsim H^{\ast }$, $\left\vert
\Delta T_{c}^{\max }\right\vert $ is at angles close to $\alpha \approx 0%
{{}^\circ}%
$.
One can infer that the strong field-direction dependence
of the superconducting onset temperature $T_{c}\left( \alpha \right) $,
 appears at high magnetic fields when the FFLO
state develops, while it is absent at low fields.

\begin{figure}[tbp]
\resizebox{0.95\columnwidth}{!}{\rotatebox{0}{
\includegraphics{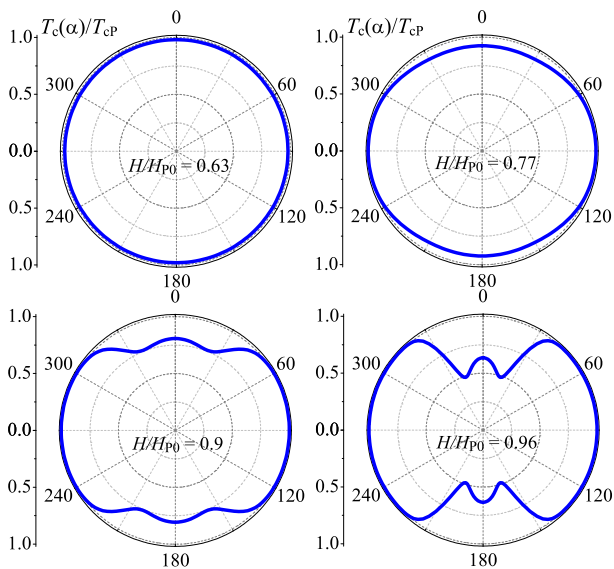}}}
\caption{Normalized superconducting transition temperature, $T_{c}\left(
\protect\alpha \right) /T_{cP}$ as a function of the angle between the
directions of the applied magnetic field and the vector $\mathbf{q}$ for
several values of $H/H_{P0}$ and for $m_{x}=m_{y}$. Here $t/T_{c0}=0.2$.}
\label{fig.02}
\end{figure}

\begin{figure}[tbp]
\resizebox{0.95\columnwidth}{!}{\rotatebox{0}{
\includegraphics{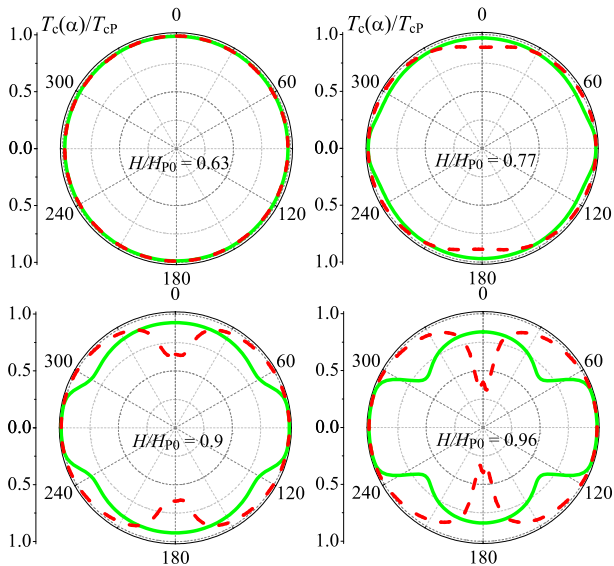}}}
\caption{The same as in Fig. \ref{fig.02} but for $m_{x}=10m_{y}$
(solid line) $m_{x}=0.1m_{y}$ (dashed line).}
\label{fig.03}
\end{figure}

The change in the anisotropy of the superconducting onset
temperature that is induced by the FFLO phase is particularly
visible in Figs. \ref{fig.02} and \ref{fig.03}, where the magnetic
field angular
dependence of the normalized superconducting transition temperature, $%
T_{c}\left( \alpha \right) /T_{cP}$, at constant modulus of the in-plane
magnetic field and $t/T_{c0}=0.2$, is plotted. In the polar plot the
direction of each point seen from the origin corresponds to the magnetic-field
direction and the distance from the origin corresponds to the
normalized critical temperature, when the orbital destructive effect is
taken into account. We show here such dependence because this type of representation
is essentially informative and was realized in the experiment.\cite{yone01}
For magnetic fields below $H^{\ast }$ and $m_{x}=m_{y}$ one can see an
expected isotropic behavior of the upper critical field. When increasing $H$
above $H^{\ast }$, a strong in-plane anisotropy of $H_{c2}$ develops, which
remains and becomes essentially pronounced at high fields. In particular,
relatively strong dips at $\alpha =\pm 18%
{{}^\circ}%
$ and $\alpha =\pm 162%
{{}^\circ}%
$ with small peaks at $0%
{{}^\circ}%
$ and $180%
{{}^\circ}%
$ develop with external magnetic field for the case of the isotropic
in-plane Fermi surface. The maximum transition temperature is for the
magnetic field orientation perpendicular to the direction of the FFLO
modulation vector along \textit{x}-axis.

\

The peculiar features of the anisotropy of the superconducting
transition onset with the orientation of the in-plane magnetic field
strongly depend on the anisotropy of the Fermi surface. For example,
comparing the curves in Fig. \ref{fig.03}, one
can notice
the totally different in-plane field-direction dependence of $%
H_{c2}$ above the tricritical point when the largest electron mass is along the
\textit{x}- (solid line) or \textit{y}-direction (dashed line).
For an in-plane mass anisotropy $m_{x}=10m_{y}$, the positions of the dips are at angles $%
\alpha =\pm 50%
{{}^\circ}%
$ and $\pm 130%
{{}^\circ}%
$, respectively. The positions of the small peaks are the same as in the
isotropic case, however they are strongly broadened. For opposite
mass anisotropy $m_{x}/m_{y}=0.1$, the dips are strengthened and
are found now at $\alpha =\pm 5%
{{}^\circ}%
$ and $\pm 175%
{{}^\circ}%
$. The fine structure at $\alpha =0%
{{}^\circ}%
,180%
{{}^\circ}%
$ is relatively sharpened. {\color{black}{Therefore the shape of the
observed $T_{c} (\alpha)$ dependence could permit to determine the
orientation of the FFLO modulation vector.}} However the common
feature of the field evolution of the in-plane upper critical field
anisotropy observed in all considered cases is that the anisotropy
becomes more pronounced with the field strength. Furthermore, the
maximum critical temperature in the FFLO regime
always corresponds to $\alpha =\pm 90%
{{}^\circ}%
$, that is $\mathbf{H}$ perpendicular to $\mathbf{q}$,
irrespective to the effective mass
ratio. This is in contrast
with the usual behavior near $T_{c0}$, where the critical temperature is
maximal for the magnetic field oriented along the lightest mass axis.

In conclusion, our results show that the FFLO modulation strongly
interferes with the orbital effect and provides the main source of
the in-plane critical field anisotropy. The superconducting
onset temperature is maximal for the field oriented perpendicular\
to the FFLO modulation vector. The change of the anisotropy of the
critical field as well as of its fine structure may give important
information about the FFLO\ state and unambiguously prove its
existence. Our calculations support the interpretation of the
experimentally observed in-plane anisotropy of the onset of
superconductivity in (TMTSF)$_{2}$ClO$_{4}$ samples as a
realization of the FFLO state with the modulation vector close to $\mathbf{b}%
^{\ast }$-axis.\cite{yone01} However, the compound
(TMTSF)$_{2}$ClO$_{4}$ is in fact in the regime $t\gtrsim T_{c0}$
($t\sim 2$-$7$ K and $T_{c0}=1.45$~K).\cite{yone01,lebe02}
{\color{black}{In this case the FFLO vector can be changed by the
orbital effect. Its orientation will result from the interplay of the
Fermi surface non-ellipticity, which favors pinning of $\mathbf{q}$
in certain direction, and the orbital effect, which prefers to orient
$\mathbf{q}$ perpendicular to $\mathbf{H}$. Nevertheless we
expect that the obtained results will be qualitatively applicable in
this case as well.}} We suggest that the predicted in-plane
anisotropy of $H_{c2}$ can be observed in experiments with $\kappa
$-(BEDT-TTF)$_{2}$Cu(NCS)$_{2}$ organic
superconductors.\cite{lort,bergk} For this salt the angle-dependent
magnetoresistance measurements\cite{singlet01} provide the estimate
of the interlayer transfer integral $t\approx 1-2$ K, which is much
smaller than $T_{c0}=9.1$ K{\color{black}{, and the orbital effect
should only slightly change the modulation vector.\cite{future}}} In
this work we have assumed \textit{s}-wave superconductivity, however
it is not an important ingredient in the present theory and we
expect similar results in the case of \textit{d}-wave pairing,
which provides an additional source of pinning for the modulation vector.%
\cite{maki}

We acknowledge the support by the European Community under a Marie Curie IEF
Action (Grant Agreement No. PIEF-GA-2009-235486-ScQSR), the French Project
\textquotedblleft SINUS\textquotedblright\ ANR-09-BLAN-0146, and thank A. S.
Mel'nikov for fruitful discussions.


\end{document}